# Beyond P-Values: Importing Quantitative Finance's Risk and Regret Metrics for AI in Learning Health Systems

**Richik Chakraborty, MS**

**Georgetown University**

**Email: rc1586@georgetown.edu**

## Abstract

The increasing deployment of artificial intelligence (AI) in clinical settings challenges foundational assumptions underlying traditional frameworks of medical evidence. Classical statistical approaches, centered on randomized controlled trials, frequentist hypothesis testing, and static confidence intervals, were designed for fixed interventions evaluated under stable conditions. In contrast, AI-driven clinical systems learn continuously, adapt their behavior over time, and operate in non-stationary environments shaped by evolving populations, practices, and feedback effects. In such systems, clinical harm arises less from average error rates than from calibration drift, rare but severe failures, and the accumulation of suboptimal decisions over time.

In this perspective, we argue that prevailing notions of statistical significance are insufficient for characterizing evidence and safety in learning health systems. Drawing on risk-theoretic concepts from quantitative finance and online decision theory, we propose reframing medical evidence for adaptive AI systems in terms of time-indexed calibration stability, bounded downside risk, and controlled cumulative regret. We emphasize that this approach does not replace randomized trials or causal inference, but complements them by addressing dimensions of risk and uncertainty that emerge only after deployment. This framework provides a principled mathematical language for evaluating AI-driven clinical systems under continual learning and offers implications for clinical practice, research design, and regulatory oversight.

## 1. Introduction

Artificial intelligence systems are increasingly integrated into clinical decision-making, supporting tasks ranging from diagnostic interpretation and prognostication to treatment recommendation and resource allocation. Unlike traditional medical interventions, many of these systems are designed to learn from new data after deployment, updating their internal representations as clinical environments evolve. Such systems are often embedded within learning health systems, where data generated during routine care are fed back into algorithmic improvement.

While this paradigm promises more responsive and personalized care, it poses fundamental challenges for established frameworks of medical evidence. Randomized controlled trials and conventional statistical inference assume that the intervention under evaluation remains fixed and that the data-generating process is sufficiently stable to support inference from historical samples. These assumptions are increasingly tenuous when applied to AI systems whose behavior changes over time and whose deployment alters the clinical environment itself.

Current approaches to evaluating clinical AI often extend traditional validation paradigms, reporting discrimination metrics, calibration curves, and statistical significance at a single point in time. However, such snapshot-based evaluations provide limited insight into how AI systems behave under distributional shift, how uncertainty evolves with continued learning, or how errors accumulate across repeated decisions. As a result, they fail to capture key dimensions of clinical risk.

In this paper, we argue that the concept of medical evidence itself requires reconsideration in the context of adaptive AI. Specifically, we propose a risk-theoretic reframing that treats evidence as an evolving assessment of calibration, downside risk, and cumulative regret over time. This perspective draws inspiration from quantitative finance and online learning, domains that have long grappled with non-stationarity, feedback loops, and the consequences of rare but catastrophic failures.

## 2. Limitations of Classical Clinical Statistics in Adaptive Systems

Classical statistical inference in medicine is built upon assumptions that are well-suited to evaluating static interventions but poorly matched to adaptive AI systems. Three assumptions are particularly salient: stationarity of the data-generating process, fixed estimands, and one-time validation.

Stationarity implies that the relationship between predictors and outcomes remains stable over time. In real-world clinical settings, however, patient populations evolve, disease prevalence shifts, clinical practices change, and new therapies are introduced. AI systems that learn continuously may amplify these dynamics by altering clinician behavior in response to their outputs, creating feedback loops that further violate stationarity.

Fixed estimands presume that the quantity of interest — such as a treatment effect or predictive performance metric — remains well-defined throughout the evaluation period. In adaptive systems, the estimand itself may evolve as the model updates, the patient population changes, or the clinical task is redefined. Under such conditions, interpreting a single confidence interval or p-value becomes conceptually problematic.

One-time validation assumes that performance assessed prior to deployment is representative of future behavior. For AI systems that continue to learn, this assumption fails by design. Post-deployment performance may diverge substantially from initial benchmarks, particularly under distributional shift or concept drift.

These limitations are not merely technical. They have direct implications for patient safety. Statistical significance at deployment offers no guarantee that an AI system will remain reliable as conditions change, nor does it quantify the potential harm arising from rare but severe failures or from the accumulation of small errors across repeated decisions.

## 3. Evidence as a Time-Indexed Process

In adaptive clinical AI systems, outcomes are more naturally represented as stochastic processes than as independent samples. Predictions, decisions, and outcomes unfold sequentially, with each action influencing future data and model updates. Under this perspective, evidence is not a static property of a model but an evolving characterization of its behavior over time.

Evaluating evidence at a single time point obscures temporal dependencies and path-dependent risks. For example, a model that performs well on average may nonetheless produce periods of sustained miscalibration following distributional shifts. Such episodes can lead to concentrated patient harm even if overall performance metrics remain acceptable.

This temporal dimension of evidence necessitates a shift in focus from point estimates to trajectories. Metrics must be capable of capturing how uncertainty, error, and risk evolve with continued use. This insight motivates the search for alternative mathematical frameworks that can represent evidence dynamically rather than statically.

## 4. Calibration Drift as a Core Clinical Risk

Among the various dimensions of model performance, calibration occupies a central role in clinical decision-making. Well-calibrated predictions allow clinicians to interpret probabilities meaningfully, supporting informed decisions about testing, treatment, and monitoring. Conversely, miscalibration undermines trust and can lead to systematic over- or under-treatment.

Importantly, calibration is not a fixed attribute of a model. Empirical studies have demonstrated that calibration can degrade over time due to changes in patient populations, disease prevalence, measurement practices, or clinical workflows (Davis et al., 2019; Davis et al., 2020; Guo et al., 2021; Nestor et al., 2019). In learning health systems, such drift may be exacerbated by feedback effects as clinicians adapt their behavior in response to algorithmic outputs.

We argue that calibration error over time should be treated as a primary unit of clinical risk. Rather than reporting a single calibration curve or summary statistic, evidence should characterize how calibration evolves, identifying periods of instability and quantifying their duration and severity.

### Box 1: Time-Indexed Calibration Error

Let $\hat{P}_t(Y=1|X)$ denote a model's predicted probability at time *t*. Time-indexed calibration error can be defined as:

$$ECE(t) = E[|\hat{P}_t(Y=1|X) - P(Y=1|X,t)|]$$

Monitoring ECE(t) over time allows detection of calibration drift even when discrimination remains stable.

---

An instructive analogy can be drawn to volatility in financial markets, where periods of relative calm are interspersed with episodes of heightened instability (Bollerslev, 1986). While health data differ fundamentally from financial time series, the conceptual parallel underscores the importance of monitoring stability over time rather than assuming persistence of historical performance.

---

## 5. Worked Clinical Examples

To ground these concepts in clinical reality, we present three problem classes where risk-theoretic evaluation reveals failures invisible to conventional metrics.

### Example 1: Calibration Drift in Sepsis Prediction Models

Consider early warning systems for sepsis deployed in hospital wards prior to the COVID-19 pandemic. Many such models demonstrated acceptable discrimination and calibration at deployment. However, during the pandemic, shifts in patient mix, respiratory failure prevalence, laboratory testing patterns, and clinical workflows substantially altered the data-generating process. Retrospective analyses have shown that several deployed sepsis models exhibited marked calibration drift during this period, systematically underestimating risk in patients with atypical inflammatory presentations (Davis et al., 2020; Finlayson et al., 2025).

Importantly, standard pre-deployment validation metrics did not flag this failure mode. Average AUC values remained relatively stable, masking the fact that predicted probabilities no longer corresponded to observed event rates. From a clinical perspective, the harm was not evenly distributed: delayed escalation occurred disproportionately among patients whose presentations fell outside pre-pandemic training distributions (Nestor et al., 2019).

Under a time-indexed evidence framework, this failure would be detected as a sustained increase in calibration error over time, triggering heightened human oversight or model retraining before widespread harm accumulated.

### Example 2: Tail Risk in ICU Mortality Prediction

ICU mortality prediction models often achieve high average performance while failing catastrophically in rare clinical scenarios, such as patients with uncommon comorbidities or atypical physiological trajectories (Ovadia et al., 2019). In these cases, a small fraction of predictions may exhibit extreme error, with predicted mortality probabilities diverging sharply from observed outcomes.

Although these events may represent less than 5% of cases, their clinical consequences are severe, influencing decisions around escalation of care, goals-of-care discussions, and resource allocation. Traditional evaluation metrics average over these failures, implicitly treating them as statistical noise.

A tail-risk perspective instead focuses explicitly on the worst-case error distribution. Bounding the expected harm among the most severe prediction failures aligns more closely with clinical safety priorities, where preventing catastrophic errors outweighs marginal improvements in mean accuracy.

### Example 3: Cumulative Regret in Oncology Decision Support

Consider a clinical decision support system that recommends surveillance versus early intervention for indolent malignancies. Even if the system is correct on average, repeated conservative recommendations can lead to delayed treatment initiation in a subset of patients whose disease progresses unexpectedly.

The resulting harm is cumulative: each delayed decision compounds future risk, leading to worse staging at diagnosis and reduced therapeutic options. Standard performance metrics evaluated at individual decision points fail to capture this accumulation (Subbaswamy & Saria, 2019).

Regret-based evaluation reframes this harm as the cumulative difference between outcomes achieved under the model's recommendations and those that would have occurred under an optimal, hindsight-informed policy. This perspective highlights long-term opportunity costs that are otherwise invisible to snapshot-based validation.

## 6. Tail Risk and the Concentration of Clinical Harm

Average performance metrics obscure the fact that clinical harm is often concentrated in rare but severe events. A small number of catastrophic errors may dominate patient outcomes, particularly in high-stakes settings such as critical care, oncology, or emergency medicine.

Risk-theoretic measures developed in quantitative finance provide a vocabulary for characterizing such phenomena (Artzner et al., 1999; Rockafellar & Uryasev, 2000). Value-at-Risk describes a threshold beyond which losses become unacceptable, while Conditional Value-at-Risk captures the expected severity of losses once that threshold is exceeded. Translated into a clinical context, these concepts correspond to bounding worst-case prediction errors or decision failures and quantifying expected harm in adverse scenarios.

## Box 2: Downside Risk and Clinical CVaR

Let *L* denote a clinically meaningful loss function (e.g., delayed intervention harm). The Conditional Value-at-Risk at level α is defined as:

$$CVaR_\alpha(L) = E[L \mid L \geq VaR_\alpha]$$

In clinical settings, $CVaR_{0.95}$ represents expected harm among the worst 5% of prediction failures, aligning evaluation with safety-critical outcomes.

We emphasize that the goal is not to import financial models wholesale into medicine, but to adopt a framework that explicitly acknowledges and quantifies downside risk. Doing so aligns more closely with clinical intuitions about safety, which prioritize avoiding rare but devastating outcomes over marginal improvements in average performance.

## Quantitative Illustration

To illustrate these concepts, consider a simulated risk prediction model deployed over 12 months with a gradual shift in outcome prevalence. While AUC remains stable around 0.83 throughout deployment, calibration error increases monotonically after month 4, rising from 0.02 to 0.12. Simultaneously, $CVaR_{0.95}$ increases from 0.08 to 0.28, indicating growing tail risk (see Figure 1).

A p-value-based validation at deployment would approve this system based on acceptable discrimination. However, continuous monitoring of ECE(t) and $CVaR_{0.95}$ would flag the model as unsafe by month 6, when both metrics exceed predetermined safety thresholds. This toy example demonstrates how risk-theoretic metrics surface clinically relevant failures missed by conventional evaluation, patterns consistent with observed degradation in deployed sepsis prediction systems during distributional shifts (Davis et al., 2020).

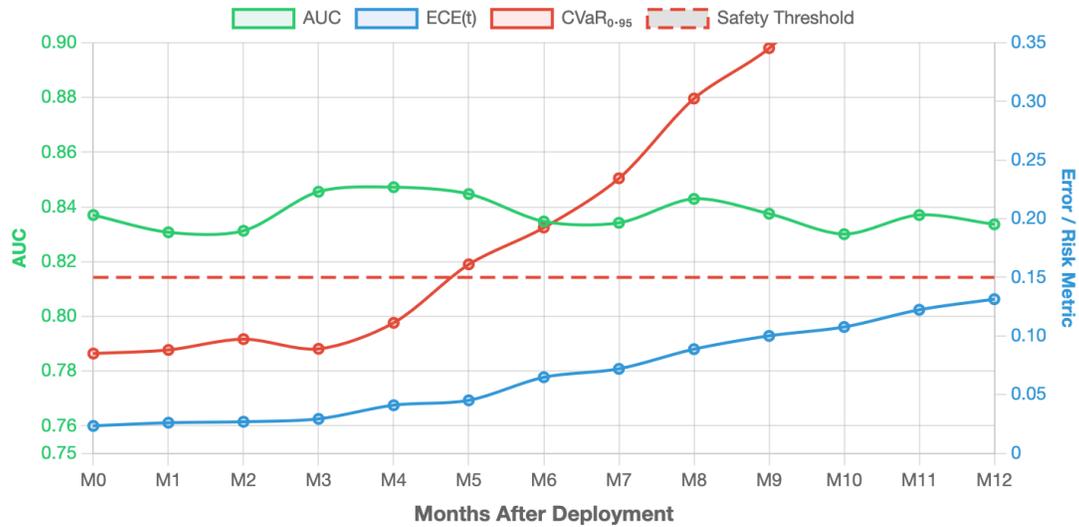

*Figure 1: Temporal evolution of risk metrics in simulated deployment. (A) AUC remains stable throughout deployment, masking underlying failures. (B) Expected Calibration Error (ECE) increases monotonically after month 4. (C) Conditional Value-at-Risk (CVaR$_{0.95}$) shows increasing tail risk, exceeding safety threshold by month 6. Red dashed line indicates predetermined safety threshold. Traditional validation at deployment would approve this system, while continuous risk monitoring would flag it as unsafe by month 6.*

## 7. Regret as Accumulated Clinical Opportunity Cost

In adaptive decision-making, the consequences of suboptimal actions accrue over time. Regret, a central concept in online learning (Lattimore & Szepesvári, 2020; Bubeck & Cesa-Bianchi, 2012), formalizes this accumulation as the difference between the outcomes achieved by a given strategy and those that would have been achieved by an optimal strategy in hindsight.

In medicine, regret corresponds to delayed initiation of effective therapies, prolonged morbidity, avoidable complications, or missed opportunities for prevention. These harms may not be apparent in cross-sectional analyses but emerge when decisions are evaluated longitudinally.

### Box 3: Cumulative Regret in Clinical Decision-Making

Over a horizon *T*, regret is defined as:

$$R(T) = \sum_{t=1}^{T} [\ell(a_t) - \ell(a_t^*)]$$

where $a_t$ is the action taken and $a_t^*$ is the optimal action in hindsight. Safety-constrained learning aims to bound R(T) while limiting exposure to unacceptable risk.

Traditional statistical frameworks lack mechanisms for quantifying such cumulative effects. By contrast, regret-based measures explicitly account for the long-run consequences of repeated decisions under uncertainty (Sutton & Barto, 2018). Incorporating regret into the definition of evidence shifts attention from isolated correctness to sustained clinical performance.

Concerns that regret minimization encourages unethical experimentation are valid and must be addressed. In clinical contexts, exploration must be constrained by asymmetric loss functions, ethical considerations, and safety requirements (Auer et al., 2002). Recent work on conservative and risk-aware learning provides potential pathways for reconciling regret minimization with clinical ethics.

## 8. A Risk-Theoretic Definition of Medical Evidence

Bringing these strands together, we propose a risk-theoretic definition of medical evidence for AI-driven clinical systems. Under this framework, evidence is characterized by four interrelated components:

1. **Posterior belief updating**, reflecting evolving uncertainty
2. **Calibration stability over time**, ensuring probabilistic reliability
3. **Bounded downside risk**, limiting worst-case patient harm
4. **Controlled cumulative regret**, constraining long-term opportunity costs

This definition preserves the strengths of probabilistic inference while addressing dimensions of risk that become salient only in adaptive, real-world deployment. It complements randomized trials and causal inference by extending the evidentiary lens beyond pre-deployment validation.

Table 1 compares this framework to conventional approaches, highlighting its unique capacity to capture temporal dynamics and concentrated harm.

*Table 1: Comparison of Evidentiary Frameworks for Clinical AI*

| Framework | Primary Metric | Temporal Awareness | Tail Risk | Cumulative Harm | Deployment Phase |
|---|---|---|---|---|---|
| RCT + p-values | Average treatment effect | None | Implicit in CI | Not measured | Pre-deployment only |
| AUC/Brier score | Discrimination/calibration | Snapshot | Averaged | Not measured | Pre-deployment + validation |
| Bayesian updating | Posterior probability | Continuous | Via posterior | Not explicit | Can be continuous |
| **Proposed: Risk-theoretic** | **ECE(t), CVaR, Regret** | **Explicit** | **Bounded** | **Monitored** | **Continuous post-deployment** |

## 9. Implications for Practice, Research, and Regulation

### Clinical Practice

This perspective suggests a shift from static validation to continuous monitoring of calibration and risk. Human oversight can be dynamically adjusted based on risk thresholds, enhancing safety without abandoning automation. When ECE(t) exceeds predetermined bounds or CVaR indicates elevated tail risk, clinical workflows can escalate to require additional human review or temporarily suspend algorithmic recommendations.

Dynamic oversight mechanisms aligned with risk thresholds offer a middle path between uncritical automation and wholesale rejection of AI assistance. By treating risk as a continuous variable rather than a binary state, clinicians can maintain appropriate situational awareness while benefiting from algorithmic support.

### Research

The framework motivates new endpoints for adaptive trials and post-deployment studies that incorporate regret and tail risk alongside traditional efficacy measures (Lu et al., 2019). Comparative evaluations should prioritize stability and safety over marginal gains in discrimination. Researchers developing adaptive AI systems should report not only AUC and calibration at deployment, but also projected bounds on calibration drift and cumulative regret under realistic deployment scenarios.

Moreover, research infrastructure must evolve to support longitudinal evaluation. Prospective cohorts with sustained follow-up become essential for characterizing temporal risk profiles, requiring sustained institutional commitment beyond typical validation study timelines.

### Regulation

Recent regulatory frameworks acknowledge these challenges. The FDA's Predetermined Change Control Plan (PCCP) and Good Machine Learning Practice (GMLP) guidance recognize that adaptive AI systems require ongoing oversight beyond initial approval (US FDA, 2021, 2023; IMDRF, 2021). However, current guidance stops short of specifying how safety and effectiveness should be quantified over time (Vokinger et al., 2021).

A risk-theoretic evidentiary framework provides operational metrics, such as calibration stability, downside risk bounds, and cumulative regret, that can be directly aligned with regulatory expectations for continuous performance monitoring. Rather than one-time approval, regulators can require ongoing demonstration that calibration, downside risk, and regret remain within acceptable bounds.

This approach enables proportionate oversight: systems with tighter risk bounds and demonstrated stability may warrant less frequent review, while those operating near safety thresholds require enhanced surveillance. The FUTUREAI international consensus guidelines similarly emphasize the need for continuous monitoring frameworks, though they do not yet specify risk-theoretic metrics (FUTUREAI Consortium, 2025).

By providing quantitative anchors for "acceptable risk," this framework offers regulators concrete decision criteria that balance innovation with patient safety — a persistent challenge in the rapidly evolving landscape of clinical AI.

## 10. Limitations and Future Directions

This framework faces important challenges, including sparse and censored clinical data, computational constraints on real-time risk estimation, and equity concerns when risk is unevenly distributed across populations.

The extension of financial risk concepts to clinical settings is not without controversy. Healthcare data exhibit fundamentally different properties than financial time series: outcomes are sparse, censored, and causally structured; populations are heterogeneous; and ethical constraints limit permissible exploration (Ashukha et al., 2020). These differences necessitate careful adaptation rather than direct transplantation of methods.

Moreover, the computational overhead of continuous risk monitoring may pose practical barriers in resource-constrained settings. Real-time calculation of time-indexed calibration error and tail risk requires streaming data infrastructure and robust statistical pipelines that may not be available in all clinical environments. Edge computing and efficient approximation algorithms may help address these barriers, but remain areas requiring methodological development.

Finally, if risk is unevenly distributed across patient subgroups, risk-based monitoring could inadvertently reinforce disparities by triggering enhanced oversight disproportionately for historically marginalized populations (Zink & Rose, 2020). Addressing these equity concerns requires explicit attention to fairness constraints within risk-theoretic frameworks, ensuring that safety mechanisms do not become instruments of bias.

Empirical validation of this framework in real-world deployments is essential. Prospective studies comparing conventional validation to risk-theoretic monitoring, ideally across multiple clinical domains and institutional settings, would provide crucial evidence for or against these proposals. Such studies should explicitly measure not only technical performance but also clinician trust, workflow integration, and health equity outcomes.

Addressing these challenges will require methodological innovation, computational investment, and careful attention to ethical implications. Nonetheless, articulating the problem space is a necessary first step toward more appropriate evidentiary standards for an era of adaptive clinical AI.

## 11. Conclusion

As AI systems become integral to clinical care, the question of what constitutes medical evidence must be revisited. In learning health systems, evidence is no longer a static declaration of efficacy but an evolving assessment of uncertainty, risk, and regret. Without a risk-theoretic framework capable of capturing these dynamics, AI will continue to outpace the mathematical tools used to govern its deployment. Reframing evidence in this way is essential for aligning innovation with patient safety.

The examples of sepsis prediction drift, ICU tail risk, and oncology regret accumulation demonstrate that these are not hypothetical concerns but observable failure modes in deployed systems. As medicine increasingly relies on systems that learn in real time, the question is no longer whether an intervention "works," but whether its uncertainty, risk, and regret remain clinically acceptable as conditions change.

This perspective does not reject randomized trials, causal inference, or traditional statistics, as these remain foundational to medical evidence. Rather, it argues for expanding the evidentiary toolkit to address challenges unique to adaptive systems operating under non-stationarity. By borrowing conceptual frameworks from domains with longer experience managing temporal risk, clinical AI can develop more mature approaches to safety and oversight.

The path forward requires collaboration across clinical medicine, machine learning, quantitative risk analysis, bioethics, and regulatory science. Only through such interdisciplinary dialogue can we develop evidentiary standards adequate to the systems we are deploying.

## Acknowledgments

The author acknowledges the use of Anthropic's Claude for assistance with mathematical structuring, and Grammarly for spell-checking assistance. All intellectual contributions, interpretations, and conclusions remain solely those of the author. No external funding was received for this work.

## Conflicts of Interest

The author declares no conflicts of interest.